\documentclass{article}

\usepackage{arxiv}

\usepackage[utf8]{inputenc} 
\usepackage[T1]{fontenc}    

\usepackage{url}            
\usepackage{booktabs}       
\usepackage{nicefrac}       
\usepackage{amsmath,amsfonts}
\usepackage{microtype}      
\usepackage{hyperref}       
\usepackage{cleveref}       
\usepackage{lipsum}         
\usepackage{natbib}
\usepackage{doi}

\usepackage{algpseudocode}
\usepackage{algorithm}
\usepackage{array}
\usepackage[caption=false,font=normalsize,labelfont=sf,textfont=sf]{subfig}
\usepackage{multirow}
\usepackage{siunitx}
\usepackage{textcomp}
\usepackage{stfloats}
\usepackage{url}
\usepackage{verbatim}
\usepackage{booktabs} 
\usepackage{graphicx}
\usepackage{cite}
\title{Rethinking Basis Path Testing: Mixed Integer Programming Approach for Test Path Set Generation}

\newif\ifuniqueAffiliation
\uniqueAffiliationtrue

\ifuniqueAffiliation 
\author{ \href{https://orcid.org/0000-0000-0000-0000}{\includegraphics[scale=0.06]{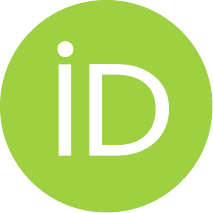}\hspace{1mm}Chao Wei} \\
	School of Computer Science and Artificial Intelligence\\
	Hubei University of Technology\\
	Wuhan 430068, China \\
	\texttt{weichao.2022@hbut.edu.cn} \\
	\And
	\href{https://orcid.org/0000-0000-0000-0000}{\includegraphics[scale=0.06]{orcid.pdf}\hspace{1mm}Xinyi Peng} \\
	School of Computer Science and Artificial Intelligence\\
	Hubei University of Technology\\
	Wuhan 430068, China \\
	\texttt{pxy\_xsyt@163.com} \\
	\And
	\href{https://orcid.org/0000-0000-0000-0000}{\includegraphics[scale=0.06]{orcid.pdf}\hspace{1mm}Yawen Yan} \\
	School of Computer Science and Artificial Intelligence\\
	Hubei University of Technology\\
	Wuhan 430068, China \\
	\texttt{102301226@hbut.edu.cn} \\
	\And
	\href{https://orcid.org/0000-0000-0000-0000}{\includegraphics[scale=0.06]{orcid.pdf}\hspace{1mm}Mao Luo}\thanks{Mao Luo is the first Corresponding author.} \\
	School of Computer Science and Artificial Intelligence\\
	Hubei University of Technology\\
	Wuhan 430068, China \\
	\texttt{luomao@hbut.edu.cn} \\
	\And
	\href{https://orcid.org/0000-0000-0000-0000}{\includegraphics[scale=0.06]{orcid.pdf}\hspace{1mm}Ting Cai}\thanks{Ting Cai is the second Corresponding author.} \\
	School of Computer Science and Artificial Intelligence\\
	Hubei University of Technology\\
	Wuhan 430068, China \\
	\texttt{caiting@hbut.edu.cn} \\
}
\else
\usepackage{authblk}

\newbox{\orcid}\sbox{\orcid}{\includegraphics[scale=0.06]{orcid.pdf}} 
\author[1]{%
	\href{https://orcid.org/0000-0000-0000-0000}{\usebox{\orcid}\hspace{1mm}Chao Wei\thanks{\texttt{weichao.2022@hbut.edu.cn}}}%
}
\author[1,2]{%
	\href{https://orcid.org/0000-0000-0000-0000}{\usebox{\orcid}\hspace{1mm}Xinyi Peng\thanks{\texttt{stariate@ee.mount-sheikh.edu}}}%
}
\affil[1]{Department of Computer Science, Cranberry-Lemon University, Pittsburgh, PA 15213}
\affil[2]{Department of Electrical Engineering, Mount-Sheikh University, Santa Narimana, Levand}
\fi

\renewcommand{\shorttitle}{\textit{arXiv} Template}

\hypersetup{
	pdftitle={A template for the arxiv style},
	pdfsubject={q-bio.NC, q-bio.QM},
	pdfauthor={Chao Wei, Xinyi Peng},
	pdfkeywords={First keyword, Second keyword, More},
}

\begin{document}
	\maketitle
	
	\begin{abstract}
		Basis path testing is a cornerstone of structural testing, yet traditional automated methods, relying on greedy graph-traversal algorithms (e.g., DFS/BFS), often generate sub-optimal paths. This structural inferiority is not a trivial issue; it directly impedes downstream testing activities by complicating automated test data generation and increasing the cognitive load for human engineers. This paper reframes basis path generation from a procedural search task into a declarative optimization problem. We introduce a Mixed Integer Programming (MIP) framework designed to produce a complete basis path set that is globally optimal in its structural simplicity. Our framework includes two complementary strategies: a Holistic MIP model that guarantees a theoretically optimal path set, and a scalable Incremental MIP strategy for large, complex topologies. The incremental approach features a multi-objective function that prioritizes path simplicity and incorporates a novelty penalty to maximize the successful generation of linearly independent paths. Empirical evaluations on both real-code and large-scale synthetic Control Flow Graphs demonstrate that our Incremental MIP strategy achieves a 100\% success rate in generating complete basis sets, while remaining computationally efficient. Our work provides a foundational method for generating a high-quality structural "scaffold" that can enhance the efficiency and effectiveness of subsequent test generation efforts.
	\end{abstract}

	\keywords{Structural Testing \and Basis Path Testing \and Test Path Generation \and Mixed Integer Programming \and Software Quality Assurance}
	
	\section{Introduction}
	Software quality assurance is a cornerstone of modern engineering, with structural (white-box) testing serving as a fundamental methodology for verifying program logic \citep{Ammann2016}. Central to this discipline is basis path testing, a technique that ensures full branch coverage by exercising a minimal set of linearly independent paths \citep{McCabe1976}. This process is fundamentally twofold: first, a set of paths—a structural "blueprint" for testing—is defined; second, concrete test data is generated to execute each of these paths. The quality of this initial path set is paramount, as it directly dictates the complexity, efficiency, and ultimate success of the subsequent automated test case generation. An improperly selected set of paths can render the test generation process computationally intractable or lead to wasted effort on infeasible execution traces. 
	
	However, a critical challenge lies in the \textit{selection} of these basis paths. While the \textit{size} of the basis set is mathematically fixed, the specific paths are not unique. Traditional procedural algorithms, such as those based on Depth-First or Bread-First Search (DFS/BFS) \citep{Watson1996, Poole1995}, operate greedily. Their localized, step-by-step search heuristic lacks a global perspective on the complex constraints of a basis set—namely, simultaneous coverage and linear independence. Consequently, especially in complex graphs with intricate loop and branch structures, these methods can easily fall into "greedy traps", making it difficult or even impossible to find a \textit{complete} set of basis paths. This fundamental limitation motivates a paradigm shift.
	
	This paper therefore argues that generating a high-quality basis path set is not a solved problem but a crucial prerequisite for effective automated testing. We reframe this task from a simple graph traversal into a declarative optimization problem: finding a complete basis set that is globally optimal in its structural simplicity. By formally encoding the properties of an ideal basis set as mathematical constraints, this approach overcomes the myopic nature of greedy selection. This shift primarily addresses the critical challenge of ensuring the completeness of the basis set—a guarantee that procedural algorithms often fail to uphold. Furthermore, it provides exceptional flexibility; by explicitly optimizing for structural simplicity, our method produces a path set that is not only complete but also composed of simple paths. Such a high-quality structural foundation is crucial, as it is widely recognized that simpler paths are more amenable to automated test data generation and easier for engineers to analyze and debug \citep{Gotlieb1998, Harman2001}.

	The application of optimization techniques to software testing is, in itself, a mature field, a domain extensively surveyed under the umbrella of Search-Based Software Testing (SBST) \citep{McMinn2004}. Seminal works have successfully employed constraint-based methods \citep{DeMillo1991} and even mixed-integer programming \citep{Lapierre1999} for the direct generation of test data to satisfy specific coverage criteria. More advanced SBST approaches, such as whole test suite generation \citep{Fraser2013}, aim to create a complete set of executable test cases in a single optimization run. However, these approaches often focus on covering individual targets (e.g., branches) rather than constructing a holistic, minimal, and high-quality set of paths that forms a complete structural basis. Our work addresses this specific, foundational gap: we are the first to formalize and solve the optimal basis path set selection problem using mathematical programming. By generating a structurally superior "scaffold" of paths, our approach serves as a powerful front-end that can enhance and simplify the task for any downstream test data generation technique.

	To this end, we present a novel Mixed Integer Programming (MIP) framework. Our contributions are fourfold:
	\begin{enumerate}
		\item We are the first to formulate basis path generation as a declarative optimization problem, shifting the focus from procedural construction to a formal definition of an optimal path set that minimizes structural complexity.
		\item We introduce a Holistic MIP model that guarantees a globally optimal basis path set, finding the simplest possible paths that collectively satisfy coverage and independence, and a robust network-flow-based constraint to ensure path integrity by eliminating subtours.
		\item To address the scalability challenges of the holistic model, we propose a scalable Incremental MIP strategy. This strategy is equipped with a multi-objective function that balances path simplicity with a "novelty penalty," a mechanism designed to conservatively explore the graph and avoid the "greedy traps" that cause traditional algorithms to fail on complex topologies.
		\item Through empirical evaluation on both real-world code and large-scale synthetic graphs, we demonstrate that our Incremental MIP approach achieves a 100\% success rate in generating complete basis sets—a level of robustness unattainable by procedural baselines—while remaining computationally efficient.
	\end{enumerate}
	
	The remainder of this paper is structured as follows: Section II details the mathematical formulation of our holistic MIP model. Section III introduces the scalable Incremental MIP strategy. Section IV presents a comprehensive empirical evaluation against baseline methods. Section V discusses the theoretical implications and limitations of our framework, and Section VI concludes with future research directions.

	\section{MIP Formulation for Basis Path Generation}
	We now present our declarative model for generating a high-quality basis path set. The problem is formulated as a Mixed Integer Programming (MIP) model, where the objective is to find a set of $k$ paths that form a basis for the Control Flow Graph (CFG) while minimizing the total path length. This section first defines the core concepts and then details the model components, including parameters, decision variables, objective function, and the constraints that formally encode the properties of a valid and optimal basis path set.
	
	\subsection{Formal Preliminaries}
	Let a program's control flow be represented by a directed graph $G=(V, E)$, where $V$ is the set of nodes (basic blocks) and $E$ is the set of edges (control transfers). A path is a sequence of connected edges from a designated source node $s$ to a sink node $t$.
	
	\subsubsection{Linearly Independent Paths}
	The concept of linear independence is central to basis path testing. A path is considered linearly independent relative to a set of existing paths if it cannot be formed by a linear combination of those paths. In our formulation, we enforce a stricter constructive condition: a path is independent if it introduces at least one edge that has not been traversed by any preceding path in the basis set.
	
	\subsubsection{Basis Path Set}
	A basis path set is a collection of linearly independent paths with two defining properties: Completeness: The set of paths must, collectively, traverse every edge in the CFG. Minimality: The size of the set is determined by the graph's cyclomatic complexity (CC), calculated as $CC = |E| - |V| + 2$.
	
	The goal is to generate exactly $k=CC$ paths that satisfy these properties while optimizing structural quality.
	
	\subsection{Sets and Parameters}
	The model is defined over the graph $G=(V, E)$:
	\begin{itemize}
		\item $V$: Set of nodes, $V = \{0, 1, \dots, n-1\}$.
		\item $E$: Set of directed edges, $E \subseteq V \times V$.
		\item $K$: Set of path indices, $K = \{1, 2, \dots, k\}$, where $k$ is the cyclomatic complexity.
		\item $s, t \in V$: The unique entry (source) and exit (sink) nodes.
		\item $M$: A sufficiently large positive constant (Big-M).
		\item $E^+(v), E^-(v)$: Sets of outgoing and incoming edges for node $v$, respectively.
	\end{itemize}
	
	\subsection{Decision Variables}
	We employ five sets of variables to capture the topology and properties of the paths. Note that the subscript $i$ denotes the $i$-th path in the set $K$.
	\begin{itemize}
		\item $x_{i,e} \in \mathbb{Z}_{\ge 0}$: Integer flow variable representing the number of times path $i$ traverses edge $e$.
		\item $y_{i,e} \in \{0, 1\}$: Binary variable indicating whether path $i$ uses edge $e$ at least once ($1$) or not ($0$).
		\item $w_{i,v} \in \{0, 1\}$: Binary variable indicating whether path $i$ visits node $v$.
		\item $f_{i,e} \in \mathbb{R}_{\ge 0}$: Continuous auxiliary flow variable used to enforce path connectivity (subtour elimination).
		\item $z_{i,e} \in \{0, 1\}$: Binary variable indicating if edge $e$ is the designated "private edge" for path $i$.
	\end{itemize}
	
	\subsection{Objective Function}
	The objective function drives the selection of the optimal basis set. While constraints ensure validity, the objective minimizes the total structural complexity, quantified by the cumulative length of all paths:
	\begin{equation}
	\min \quad Z = \sum_{i \in K} \sum_{e \in E} x_{i,e}
	\label{eq:objective}
	\end{equation}
	Eq. \ref{eq:objective} serves two critical purposes: 1) Suppression of redundant cycles. In a CFG with loops, infinite valid paths exist. By assigning a unit cost to every edge traversal, the solver is mathematically forced to prune unnecessary loop iterations. A cycle is traversed only if it is strictly required to reach a specific node or to cover a specific edge for linear independence. 2) Cognitive simplicity. Shorter paths correspond to simpler execution traces. By seeking the global minimum of $Z$, the model inherently prefers the "canonical" paths, reducing the cognitive load for human testers and the computational cost of execution.
	
	\subsection{Constraints}
	The constraints are grouped into four categories: path validity, connectivity, variable consistency, and basis set properties.
	
	\subsubsection{Path Flow Conservation}
	Standard network flow constraints ensure that the decision variables $x_{i,e}$ form a continuous route from $s$ to $t$:
	\begin{equation}
	\sum_{e \in E^+(v)} x_{i,e} - \sum_{e \in E^-(v)} x_{i,e} =
	\begin{cases}
	1 & \text{if } v = s \\
	-1 & \text{if } v = t \\
	0 & \text{otherwise}
	\end{cases}
	\forall i \in K, \forall v \in V \label{eq:main_flow}
	\end{equation}
	Eq.~\eqref{eq:main_flow} enforces Kirchhoff's law \citep{Ahuja1993}: flow is generated at $s$, conserved at intermediate nodes, and absorbed at $t$.

	\subsubsection{Path Connectivity (Subtour Elimination)}
	Standard flow conservation (Eq.~\eqref{eq:main_flow}) allows for a valid path from $s$ to $t$ to coexist with disjoint cycles (subtours) that are disconnected from the main path. To strictly prohibit such disconnected components, we implement a single-source auxiliary flow formulation. We treat $s$ as a source of "commodity flow" required by every visited node:
	
	\begin{align}
		\sum_{e \in E^+(v)} f_{i,e} - \sum_{e \in E^-(v)} f_{i,e} &=
		\begin{cases}
			\sum\limits_{u \in V \setminus \{s\}} w_{i,u} & \text{if } v = s \\
			-w_{i,v} & \text{if } v \neq s
		\end{cases} \nonumber \\
		& \quad \forall i \in K, \forall v \in V \label{eq:aux_flow} \\
		f_{i,e} &\le (|V| - 1) \cdot y_{i,e} \nonumber \\
		& \quad \forall i \in K, \forall e \in E \label{eq:aux_capacity}
	\end{align}
	
	To illustrate the necessity and mechanism of these constraints, consider a scenario where the solver attempts to form a solution comprising a valid path $P: s \to \dots \to t$ and a disjoint cycle $C: u \to v \to u$ (where $u, v$ are not reachable from $s$ via active edges). Under Eq.~\eqref{eq:main_flow} alone, this solution is valid because flow is conserved at $u$ and $v$ (inflow equals outflow). However, under Eq.~\eqref{eq:aux_flow}, since nodes $u$ and $v$ are active ($w_{i,u}=w_{i,v}=1$), they each demand 1 unit of auxiliary flow from $s$. Eq.~\eqref{eq:aux_capacity} dictates that auxiliary flow $f$ can only pass through edges selected in the path ($y_{i,e}=1$). Since cycle $C$ is disconnected from $s$, there exists no continuous chain of active edges from $s$ to $u$. Consequently, the flow reaching $u$ is zero ($\sum f_{in} = 0$). This contradicts Eq.~\eqref{eq:aux_flow} for node $u$, which requires a net flow of $-1$. Thus, the model declares this solution infeasible, forcing the elimination of the subtour. In essence, Eq.~\eqref{eq:aux_flow} ensures that every active node lies on a continuous path originating from $s$, mathematically guaranteeing a single connected component.
	
	\subsubsection{Variable Linking and Consistency}
	Coupling constraints synchronize the integer flow ($x$), binary usage ($y$), and node activation ($w$) variables:
	\begin{align}
		x_{i,e} &\le M \cdot y_{i,e} && \forall i \in K, \forall e \in E \label{eq:link_xy_1} \\
		x_{i,e} &\ge y_{i,e} && \forall i \in K, \forall e \in E \label{eq:link_xy_2} \\
		w_{i,s} &= 1 && \forall i \in K \label{eq:link_w_s}
	\end{align}
	\begin{align}
		\sum_{e \in \delta(v)} y_{i,e} &\le M \cdot w_{i,v} && \forall i \in K, \forall v \in V \setminus \{s\} \label{eq:link_wy_1} \\
		w_{i,v} &\le \sum_{e \in \delta(v)} y_{i,e} && \forall i \in K, \forall v \in V \setminus \{s\} \label{eq:link_wy_2}
	\end{align}
	where $\delta(v) = E^+(v) \cup E^-(v)$. These ensure that $y_{i,e}=1 \iff x_{i,e} \ge 1$, and $w_{i,v}=1$ if and only if an incident edge is used.
	
	\subsubsection{Basis Set Properties: Coverage and Independence}
	Finally, we enforce the collective properties of the set $K$.
	\begin{equation}
	\sum_{i \in K} y_{i,e} \ge 1 \quad \forall e \in E \label{eq:coverage}
	\end{equation}
	Eq.~\eqref{eq:coverage} ensures Completeness: every edge is covered by at least one path.
	
	To enforce linear Independence, we employ an asymmetric "private edge" constraint. This forces the incidence matrix of the path set to have a triangular structure, which is a sufficient condition for independence:
	\begin{align}
		\sum_{e \in E} z_{i,e} &\ge 1 && \forall i \in K \label{eq:private_sum} \\
		z_{i,e} &\le y_{i,e} && \forall i \in K, \forall e \in E \label{eq:private_link} \\
		\sum_{j=1}^{i-1} y_{j,e} &\le (1 - z_{i,e}) \cdot M && \forall i \in K \setminus \{1\}, \forall e \in E \label{eq:private_def}
	\end{align}
	Eq.~\eqref{eq:private_sum} requires every path $i$ to identify at least one private edge ($z_{i,e}=1$). Eq.~\eqref{eq:private_link} ensures a private edge must be part of the path. Crucially, Eq.~\eqref{eq:private_def} dictates that if edge $e$ is private for path $i$ ($z_{i,e}=1$), then no preceding path $j < i$ could have used it ($\sum y_{j,e} = 0$). This ensures that each path $i$ introduces structural novelty relative to $\{1, \dots, i-1\}$, mathematically guaranteeing the independence of the entire set.
	\section{An Incremental MIP Approach for Basis Path Generation}
	While the holistic MIP model presented in Section II guarantees a globally optimal basis set, it requires solving a single, large-scale optimization problem involving $k$ simultaneous paths. As $k$ increases, the coupling constraints between paths lead to a combinatorial explosion in the solution space. To address this scalability challenge, we propose an Incremental MIP Strategy. This approach decomposes the global problem into a sequence of $k$ independent, smaller MIP models, where each iteration generates exactly one new path that is linearly independent of the previously accumulated set.
	
	\subsection{Iterative Model Formulation}
	The core principle is to construct the basis set $\mathcal{B} = \{P_1, P_2, \dots, P_k\}$ sequentially. In each iteration $i$ (where $1 \le i \le k$), we formulate a MIP model to generate path $P_i$. Crucially, this model relies on the state of the Covered Edge Set, denoted as $E_{cov}^{(i-1)}$, which contains the union of all edges traversed by paths $P_1$ through $P_{i-1}$.
	
	For the $i$-th iteration, we define the decision variables for a single path (dropping the path index $i$ for brevity):
	\begin{itemize}
		\item $x_e \in \mathbb{Z}_{\ge 0}$: Flow on edge $e$.
		\item $y_e \in \{0, 1\}$: Usage of edge $e$.
		\item $w_v \in \{0, 1\}$: Visitation of node $v$.
		\item $f_e \in \mathbb{R}_{\ge 0}$: Auxiliary flow for connectivity.
	\end{itemize}
	
	\begin{figure}[t]
		\centering
		\includegraphics[width=1.2in]{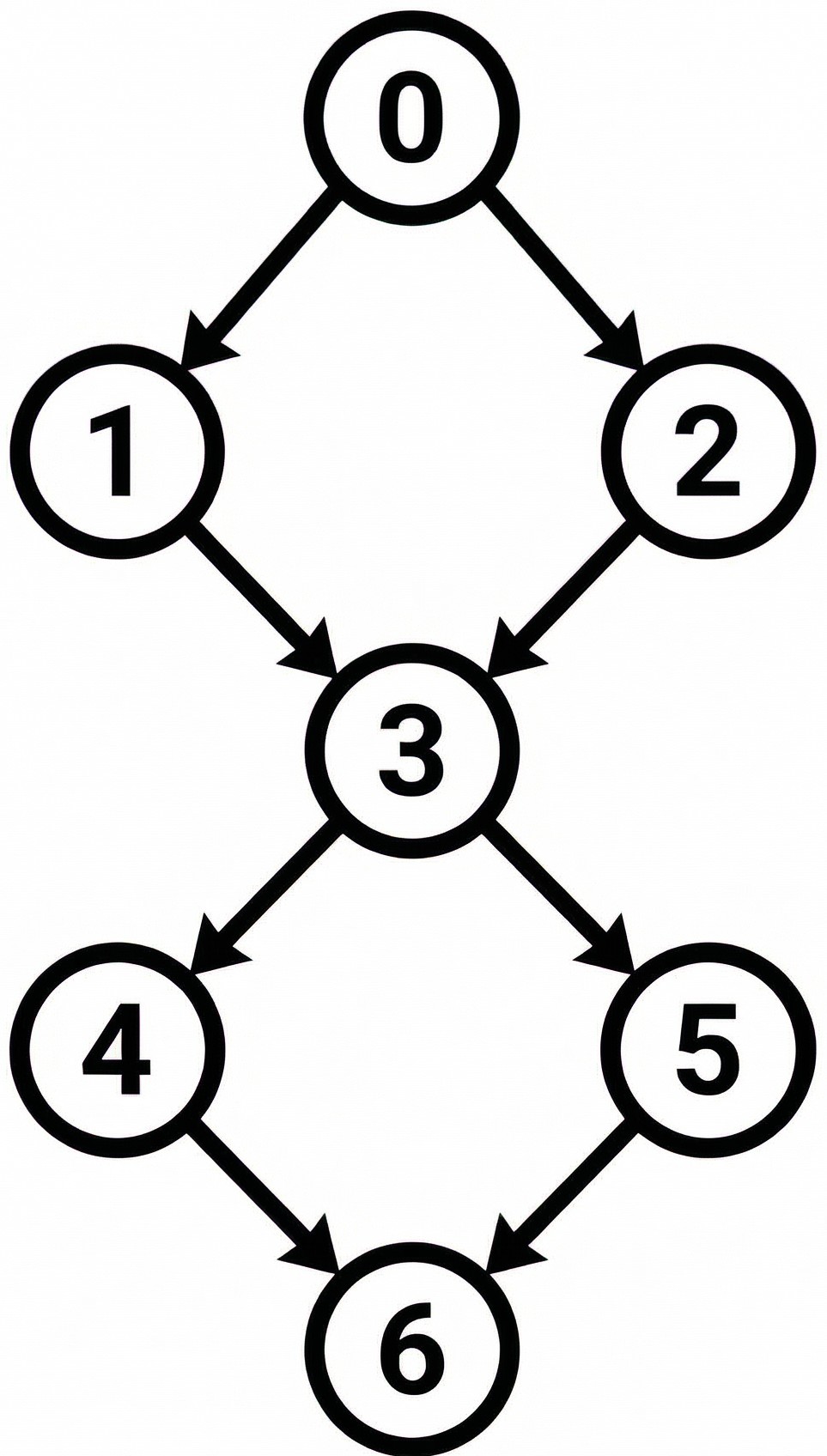} 
		\caption{The "Double Diamond" CFG structure. With $|V|=7$ nodes and $|E|=8$ edges, the cyclomatic complexity is $k = 8 - 7 + 2 = 3$. This topology perfectly illustrates the "Greedy Trap": a naive algorithm might select two paths (e.g., $0 \to 1 \to 3 \to 4 \to 6$ and $0 \to 2 \to 3 \to 5 \to 6$) that cover all edges. This makes it impossible to generate a third, linearly independent path, causing the algorithm to fail in producing a complete basis set.}
		\label{fig:double_diamond}
	\end{figure}
	
	\subsubsection{Objective Function: Dual Optimization of Length and Novelty}
	
	A naive approach would simply minimize path length ($\sum x_e$) at each step. However, as illustrated by the "Double Diamond" structure in Fig.~\ref{fig:double_diamond}, purely greedy length minimization can lead to a "basis deadlock." Consider the CFG in Fig.~\ref{fig:double_diamond}, which has a cyclomatic complexity of $k=3$. A greedy algorithm might first select the path $0 \to 1 \to 3 \to 4 \to 6$. For its second path, to maximize novelty or simply find another short path, it might select $0 \to 2 \to 3 \to 5 \to 6$. While these two paths are valid and independent, their union covers all 8 edges of the graph. This premature consumption of all available edges creates a deadlock: it is now impossible for the algorithm to generate a third path that introduces a new edge, as required by the linear independence constraint. Consequently, the greedy approach fails to generate a complete basis set of size 3.

	To mitigate this, we employ a multi-objective function that simultaneously minimizes structural complexity and encourages conservative exploration, resulting in the following objective:
	
	\begin{equation}
	\min \quad Z_i = \underbrace{\sum_{e \in E} x_{e}}_{\text{Path Length}} + \underbrace{\sum_{e \in E \setminus E_{cov}^{(i-1)}} y_{e}}_{\text{Novelty Penalty}}
	\label{eq:incremental_objective}
	\end{equation}
	
	This formulation assigns a "dual cost" to structural novelty. Every edge traversal incurs a unit cost (representing length), but traversing an edge that has not yet been covered incurs an \textit{additional} unit penalty. This mechanism creates a robust heuristic: the solver is mathematically incentivized to prioritize paths that reuse existing structural patterns ($E_{cov}^{(i-1)}$) and is deterred from "prematurely" consuming easy-to-reach new edges unless necessary.
	
	\subsubsection{Constraints}
	The model for iteration $i$ inherits the structural validity constraints from the holistic model, adapted for a single path:
	\begin{itemize}
		\item \textbf{Flow Conservation:} Eq.~\eqref{eq:main_flow} applied to $x_e$.
		\item \textbf{Connectivity:} Eqs.~\eqref{eq:aux_flow}--\eqref{eq:aux_capacity} applied to $f_e, w_v, y_e$.
		\item \textbf{Consistency:} Eqs.~\eqref{eq:link_xy_1}--\eqref{eq:link_wy_2} applied to $x_e, y_e, w_v$.
	\end{itemize}
	
	The distinguishing constraint for the incremental approach is the linear independence constraint:
	\begin{equation}
	\sum_{e \in E \setminus E_{cov}^{(i-1)}} y_{e} \ge 1
	\label{eq:incremental_independence}
	\end{equation}
	Constraint~\eqref{eq:incremental_independence} mandates that the new path must traverse at least one edge that is \textit{not} currently in the covered set. This strictly ensures that the incidence vector of the new path is linearly independent of the vectors of all previous paths. For the first iteration ($i=1$), $E_{cov}^{(0)} = \emptyset$, and the constraint simply requires the path to not be empty.
	
	\subsection{Algorithmic Process}
	The formal procedure for the Incremental MIP strategy is outlined in Algorithm~\ref{alg:incremental_mip}. This algorithm systematically constructs the basis set $\mathcal{B}$ by solving a sequence of $k$ optimization problems.
	
	\begin{algorithm}[h]
		\caption{Incremental MIP-based Basis Path Generation}
		\label{alg:incremental_mip}
		\begin{algorithmic}[1]
			\State \textbf{Input:} Control Flow Graph $G=(V, E)$, Source node $s$, Sink node $t$
			\State \textbf{Output:} Basis Path Set $\mathcal{B} = \{P_1, P_2, \dots, P_k\}$
			
			\State \textit{// Phase 1: Initialization}
			\State Calculate Cyclomatic Complexity, $k \gets |E| - |V| + 2$ 
			\State Initialize global covered edge set, $E_{cov} \gets \emptyset$ 
			\State Initialize empty basis set, $\mathcal{B} \gets \emptyset$ 
			
			\State
			\State \textit{// Phase 2: Iterative Construction}
			\For{$i \gets 1$ to $k$}
			\State \textbf{Formulate MIP Model $\mathcal{M}_i$:}
			\State \quad \textbf{Objective:}
			\State \quad \quad Minimize $Z_i = \sum_{e \in E} x_{e} + \sum_{e \in E \setminus E_{cov}} y_{e}$
			\State \quad \textbf{Subject to:}
			\State \quad \quad 1. Flow Conservation (Eq.~\ref{eq:main_flow})
			\State \quad \quad 2. Subtour Elimination (Eqs.~\ref{eq:aux_flow}--\ref{eq:aux_capacity})
			\State \quad \quad 3. Consistency Constraints (Eqs.~\ref{eq:link_xy_1}--\ref{eq:link_wy_2})
			\State \quad \quad 4. Linear Independence: $\sum_{e \in E \setminus E_{cov}} y_{e} \ge 1$
			
			\State $(x^*, y^*) \gets \text{SOLVER}(\mathcal{M}_i)$ 
			
			\If{No Solution Found}
			\State \textbf{Error:} "Topology Infeasible or Timeout"
			\State \textbf{Break}
			\EndIf
			\State
			\State \textit{// Phase 3: Update State}
			\State $P_i \gets \text{ExtractPathFromFlow}(x^*)$
			\State Identify new edges, $E_{new} \gets \{e \in E \mid y_e^* = 1\}$ 
			\State Update historical coverage, $E_{cov} \gets E_{cov} \cup E_{new}$ 
			\State Update Basis Path Set, $\mathcal{B} \gets \mathcal{B} \cup \{P_i\}$
			\EndFor
			
			\State
			\State \textbf{Return} $\mathcal{B}$
		\end{algorithmic}
	\end{algorithm}
	
	As shown in Algorithm~\ref{alg:incremental_mip}, the process begins by calculating the target basis size $k$ and initializing the required sets (Lines 4-6). The core loop (Lines 9-29) iterates exactly $k$ times. In each iteration, the objective function dynamically adapts its structure based on the current covered edge set $E_{cov}$. The novelty penalty mechanism is critical: by making the consumption of new edges costly, it guides the solver to satisfy the linear independence constraint in the most conservative way possible. This strategic preservation of uncovered edges significantly enhances the probability of successfully generating the remaining linearly independent paths in subsequent iterations. Finally, the newly found path is added to the basis set, and the coverage history is updated for the next cycle.
	
	\subsection{Trade-off Analysis}
	The Incremental strategy presents a distinct trade-off between computational scalability and theoretical global optimality, governed by the novelty-driven objective function:
	
	\begin{itemize}
		\item Computational Scalability: By decomposing the monolithic optimization problem into $k$ sequential, smaller-scale MIP instances, the Incremental strategy effectively linearizes the complexity relative to the basis size. This decoupling ensures that the resource consumption for each sub-problem remains constant and manageable, allowing the method to scale to large-scale CFGs where the Holistic model would typically encounter memory exhaustion or timeout constraints due to the combinatorial explosion of simultaneous path generation.
		
		\item Robustness via Novelty Penalty: While the sequential approach lacks the "foresight" to guarantee the absolute completeness of the generated basis path set (Global Optimality), the proposed novelty penalty mechanism in the objective function effectively mitigates the limitations of greedy heuristics. By assigning an additional cost to the traversal of uncovered edges, the model is mathematically incentivized to be "structurally conservative"—reusing existing paths whenever possible and consuming new edges only when strictly necessary for linear independence. This strategy prevents the premature exhaustion of easy-to-reach edges, thereby maximizing the algorithm's ability to successfully discover the remaining independent paths in complex topologies and ensuring a 100\% generation success rate.
	\end{itemize}
	
	\section{Case Study and Empirical Evaluation}
	\label{sec:evaluation}
	
	In order to fully assess the effectiveness, efficiency, and robustness of the proposed Mixed Integer Programming (MIP) algorithms, we carried out a systematic empirical study. The assessment is meant to make comparisons of our approaches against a traditional procedural setting and to evaluate the trade-offs of our Holistic and Incremental MIP formulations.
	
	\subsection{Experimental Setup}
	\subsubsection{Environment and Baseline}
	Environment and Baseline: All algorithms were implemented in Python 3.8, with the MIP models based on the IBM ILOG CPLEX Optimizer (v12.10.0.0) as the solver. A standard Breadth-First Search (BFS) based procedural algorithm, adapted from the principles in \citep{Poole1995}, served as the baseline for comparison. All experiments were performed on a workstation with an Intel Core i7-11700F processor (2.50 GHz) and 16GB of RAM, with a 3600-second timeout for each run.
	
	\subsubsection{MIP Configurations}
	We evaluated two distinct incremental MIP strategies to investigate the impact of the objective function. Incr. MIP1: An incremental approach that greedily minimizes only the path length at each step. Incr. MIP2: The other incremental approach, which employs a multi-objective function to minimize a sum of the path length and the number of new edges introduced.
	
	\subsubsection{Dataset}
	The evaluation utilized two distinct categories of Control Flow Graphs to ensure both practical applicability and stress-testing capabilities:
	
	\begin{itemize}
		\item Real-code Dataset: To evaluate the approach on actual software artifacts, we curated a dataset of 50 Python functions. These functions cover a wide range of logic patterns, including matrix operations, string processing, and multi-condition validations. The Control Flow Graphs were generated by parsing the source code using the \texttt{staticfg} tool \citep{staticfg}, which constructs the graph topology directly from the Python Abstract Syntax Tree (AST) \citep{Cormen2009}. The CC of these real-code instances ranges from 1 to 8, representing typical complexity levels found in unit testing scenarios.
		\item Synthetic Large-scale Dataset: For quantitative scalability analysis, we employed a large-scale dataset of randomly generated CFGs. This dataset is divided into three groups with controlled complexities of $(CC=10, |V|=9)$, $(CC=50, |V|=30)$, and $(CC=100, |V|=50)$, designed to test the algorithms beyond typical human-written code complexity.
	\end{itemize}
	
	\subsection{Qualitative Comparison of Path Sets}
\begin{figure}[h]
		\centering
		\includegraphics[width=2.2in]{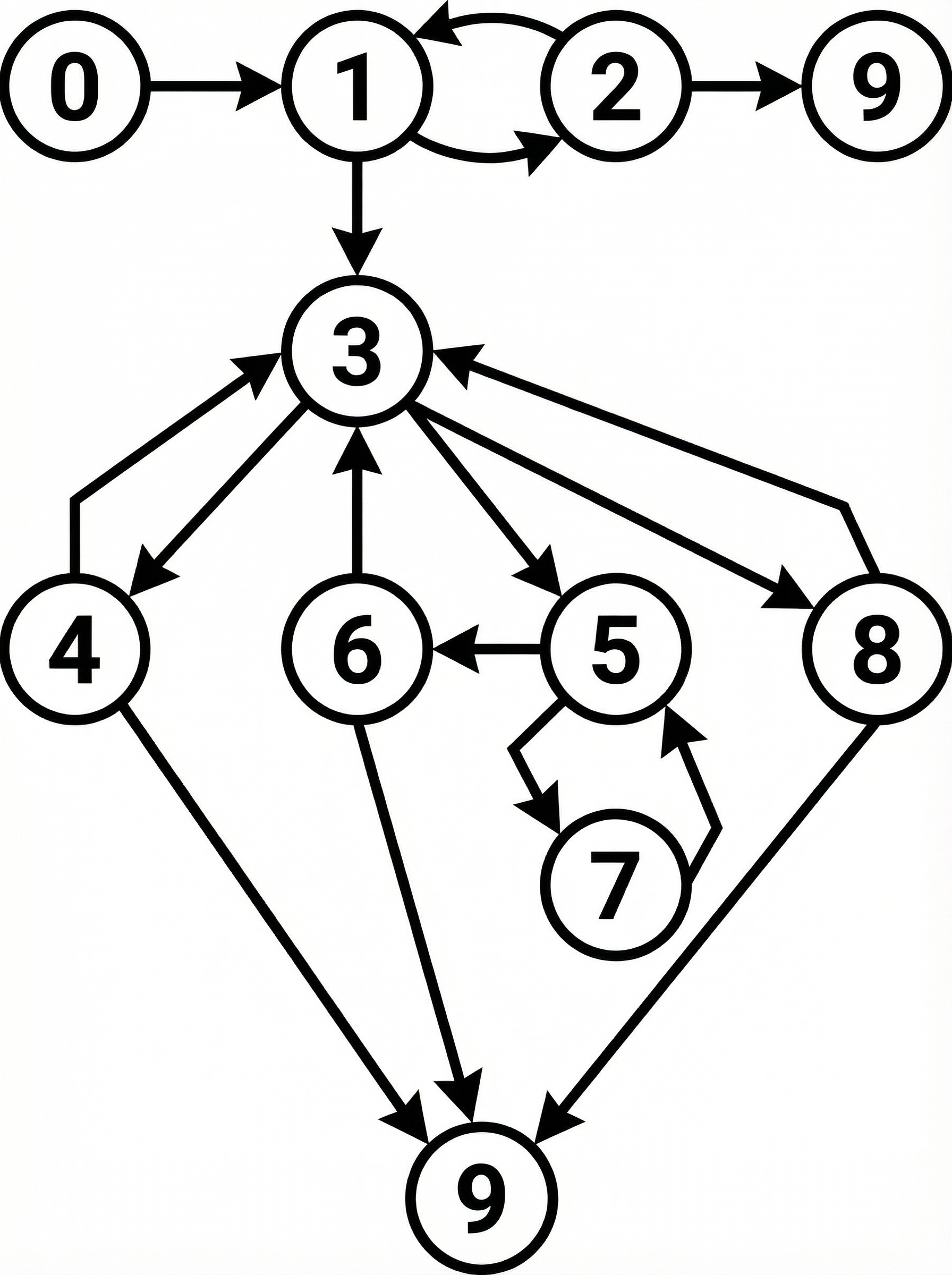} 
		\caption{The illustrative Control Flow Graph (CFG) used for the qualitative analysis. This graph features a single entry (Node 0) and exit (Node 9) and contains multiple loops and branches, resulting in a cyclomatic complexity ($k$) of 9. It serves as the input for generating the basis path sets compared in Table~\ref{tab:integrated_paths}.}
		\label{fig:robustness} 
	\end{figure}

	To visualize the generation capabilities and validate the necessity of our mathematical constraints, we applied the Baseline BFS, the Holistic MIP strategies, and the Incremental MIP approaches to the illustrative Control Flow Graph (CFG) shown in Fig.~\ref{fig:robustness}. The resulting path sets are detailed in Table~\ref{tab:integrated_paths}.
	
	A critical comparison is presented between Holistic MIP1 (the optimization model without the auxiliary flow connectivity constraints) and Holistic MIP2 (the complete model). As observed in Table~\ref{tab:integrated_paths}, Holistic MIP1 fails to produce valid executable traces. For instance, in Path 4 of Holistic MIP1 ($0 \to 1 \to 2 \to 9;\quad 3 \to 4 \to 3$), the solver satisfies the flow conservation constraints by generating a valid path from Source to Sink ($0 \to \dots \to 9$) alongside a disjoint, isolated cycle ($3 \to 4 \to 3$). This phenomenon occurs because, without the strict connectivity enforcement provided by our network-flow-based auxiliary constraints, the solver exploits disjoint loops to satisfy node flow balance at a lower total "cost" (length) than a single continuous path.
	
	In contrast, Holistic MIP2 incorporates the proposed subtour elimination constraints. As shown in the table, it successfully suppresses these disconnected components, forcing the solver to construct valid, continuous paths (e.g., Path 4 becomes $0 \to 1 \to 3 \to 5 \to 6 \to 9$) that maintain the global optimality of the basis set. Furthermore, the Incremental MIP2 (Novelty-Driven) and the Baseline BFS also produce valid continuous paths. 
	
	\subsection{Performance on Real Code}
	Table~\ref{tab:realdata} presents the performance metrics in terms of Success Rate (percentage of runs generating exactly $k=CC$ linearly independent paths), Path Coverage Rate (percentage of generating  correct linearly independent paths), and average Execution Time on the Real-code Dataset. For these practical instances (CC $\in [1, 8]$), the MIP-based approaches demonstrated absolute reliability, achieving a 100\% success rate and 100\% path coverage. While the Baseline BFS performed adequately with a 90\% success rate, it still failed to generate complete basis sets for 5 specific functions containing specific loop structures. In terms of efficiency, the Incremental MIP strategies solved these instances in approximately 0.03 seconds, proving that the overhead of the optimization model is negligible for standard software units.
	
	\subsection{Scalability and Robustness Evaluation}
	To assess how well the proposed methods perform or scale with increasing topological complexity, we performed a large-scale quantitative experiment. For each complexity group defined in the dataset, we set 1,000 independent runs. The aggregated results are summarized in Table~\ref{tab:scalability} in terms of Success Rate, Path Coverage Rate, and average Execution Time. 
	
	The results show a considerable distinction in terms of robustness between procedural and optimization-based methods. The Baseline BFS, while computationally negligible in cost (near 0.00s), demonstrates fundamental unreliability. Even on the simplest graphs ($CC=10$), its success rate is only 27.8\%, which further deteriorates to 12.3\% as complexity rises to $CC=100$. Although BFS achieves high edge coverage ($>97\%$), it consistently fails to identify the paths required to satisfy the full basis cardinality, confirming that greedy graph traversal is insufficient for rigorous basis path generation.
	
	The Holistic MIP model exhibits the expected trade-off between optimality and tractability. While it achieves perfect success on small topologies ($CC=10$), it hits computational limits on larger instances. The missing execution time data (marked as $-$) for complexities of 50 and 100 indicates that the model consistently exceeded the 3600-second timeout, rendering it impractical for large-scale automated testing despite its theoretical guarantees.
	
	The most significant finding lies in the comparison between the two incremental strategies. The Incr. MIP1 (Greedy) strategy suffers from a severe "Greedy Trap." As complexity increases, its success rate plummets from 97.5\% to a mere 14.7\%. By myopically minimizing path length, it prematurely consumes critical edges, making it mathematically impossible to form subsequent independent paths without violating structural constraints.
	
	In contrast, Incr. MIP2 (Novelty-Driven) strategy demonstrates superior robustness, maintaining a 100\% Success Rate across all complexity levels. By balancing path length with a penalty for structural novelty, it effectively navigates complex control flows without reaching deadlocks. Furthermore, it remains computationally efficient, solving the most complex instances ($CC=100$) in approximately 17.7 seconds. This result empirically validates that the multi-objective Incremental MIP is the only evaluated method capable of guaranteeing both the completeness of the basis set and practical scalability for complex software systems.

	\begin{table}[htbp]
		\centering
		\caption{Qualitative Comparison of Basis Path Sets for the Illustrative CFG ($k=9$). }
		\label{tab:integrated_paths}
		\setlength{\tabcolsep}{4pt}
		
		\begin{tabular}{@{}lcl@{}} 
			\toprule
			\textbf{Method} & \textbf{Path} & \textbf{Path Trace} \\
			\midrule
			
			\multirow{9}{*}{BFS}
			& 1 & $0 \to 1 \to 2 \to 9$  \\
			& 2 & $0 \to 1 \to 3 \to 4 \to 9$  \\
			& 3 & $0 \to 1 \to 2 \to 1 \to 2 \to 9$  \\
			& 4 & $0 \to 1 \to 3 \to 5 \to 6 \to 9$  \\
			& 5 & $0 \to 1 \to 3 \to 8 \to 9$  \\
			& 6 & $0 \to 1 \to 3 \to 4 \to 3 \to 4 \to 9$  \\
			& 7 & $0 \to 1 \to 3 \to 5 \to 7 \to 5 \to 6 \to 9$  \\
			& 8 & $0 \to 1 \to 3 \to 5 \to 6 \to 3 \to 4 \to 9$  \\
			& 9 & $0 \to 1 \to 3 \to 8 \to 3 \to 4 \to 9$  \\
			\midrule
			
			\multirow{9}{*}{Holistic MIP1} 
			& 1 & $0 \to 1 \to 3 \to 8 \to 9$  \\
			& 2 & $0 \to 1 \to 2 \to 9$ \\
			& 3 & $0 \to 1 \to 3 \to 4 \to 9$  \\
			& 4 & $0 \to 1 \to 2 \to 9;\quad 3 \to 4 \to 3$ \\
			& 5 & $0 \to 1 \to 2 \to 9;\quad 5 \to 7 \to 5$ \\
			& 6 & $0 \to 1 \to 2 \to 9;\quad 3 \to 5 \to 6 \to 3$ \\
			& 7 & $0 \to 1 \to 2 \to 1 \to 2 \to 9$ \\
			& 8 & $0 \to 1 \to 2 \to 9;\quad 3 \to 8 \to 3$ \\
			& 9 &$0 \to 1 \to 3 \to 5 \to 6 \to 9$  \\
			\midrule
			\multirow{9}{*}{Holistic MIP2} 
			& 1 & $0 \to 1 \to 2 \to 9$ \\
			& 2 & $0 \to 1 \to 3 \to 4 \to 9$  \\
			& 3 & $0 \to 1 \to 3 \to 8 \to 9$  \\
			& 4 &$0 \to 1 \to 3 \to 5 \to 6 \to 9$  \\
			& 5 & $0 \to 1 \to 2 \to 1 \to 3 \to 4 \to 9$  \\
			& 6 & $0 \to 1 \to 3 \to 4 \to 3 \to 8 \to 9$  \\
			& 7 & $0 \to 1 \to 3 \to 8 \to 3 \to 4 \to 9$  \\
			& 8 & $0 \to 1 \to 3 \to 5 \to 6 \to 3 \to 4 \to 9$  \\
			& 9 & $0 \to 1 \to 3 \to 5 \to 7 \to 5 \to 6 \to 9$  \\
			\midrule
			\multirow{9}{*}{Incr. MIP1}
			& 1 & $0 \to 1 \to 2 \to 9$ \\
			& 2 & $0 \to 1 \to 3 \to 4 \to 9$ \\
			& 3 & $0 \to 1 \to 3 \to 8 \to 9$ \\
			& 4 & $0 \to 1 \to 3 \to 5 \to 6 \to 9$ \\
			& 5 & $0 \to 1 \to 2 \to 1 \to 2 \to 9$ \\
			& 6 & $0 \to 1 \to 3 \to 8 \to 3 \to 4 \to 9$  \\
			& 7 & $0 \to 1 \to 3 \to 4 \to 3 \to 4 \to 9$ \\
			& 8 & $0 \to 1 \to 3 \to 5 \to 7 \to 5 \to 6 \to 9$ \\
			& 9 & $0 \to 1 \to 3 \to 5 \to 6 \to 3 \to 4 \to 9$ \\
			
			\midrule
			\multirow{9}{*}{Incr. MIP2}
			& 1 & $0 \to 1 \to 2 \to 9$ \\
			& 2 & $0 \to 1 \to 2 \to 1 \to 2 \to 9$ \\
			& 3 & $0 \to 1 \to 3 \to 4 \to 9$ \\
			& 4 & $0 \to 1 \to 3 \to 8 \to 9$ \\
			& 5 & $0 \to 1 \to 3 \to 4 \to 3 \to 4 \to 9$ \\
			& 6 & $0 \to 1 \to 3 \to 8 \to 3 \to 4 \to 9$  \\
			& 7 & $0 \to 1 \to 3 \to 5 \to 6 \to 9$ \\
			& 8 & $0 \to 1 \to 3 \to 5 \to 6 \to 3 \to 4 \to 9$ \\
			& 9 & $0 \to 1 \to 3 \to 5 \to 7 \to 5 \to 6 \to 9$ \\
			
			\bottomrule
		\end{tabular}
	\end{table}
	
	\begin{table}[h]
		\centering
		\caption{Performance Comparison on the Real-code Dataset (50 Python Functions). The dataset represents typical unit-level complexity ($k \in [1, 8]$) generated via AST analysis.}
		\label{tab:realdata}
		\setlength{\tabcolsep}{3.5pt} 
		
		\begin{tabular}{@{}cccc@{}} 
			\toprule
			\textbf{Method} & 
			\textbf{\begin{tabular}{@{}c@{}}Success \\ Rate (\%)\end{tabular}} & 
			\textbf{\begin{tabular}{@{}c@{}}Coverage \\ Rate (\%)\end{tabular}} & 
			\textbf{\begin{tabular}{@{}c@{}}Time \\ (s)\end{tabular}}  \\ 
			\midrule
			
			BFS & 90 & 96.72 & \textbf{0.0001}  \\ 
			Holistic MIP & 100 & 100 & 0.1437  \\ 
			Incr. MIP1 & 100 & 100 & 0.0318 \\ 
			Incr. MIP2 & 100 & 100 & 0.0312 \\ 
			\bottomrule          
		\end{tabular}
	\end{table}

	\begin{table}[h]
		\centering
		\caption{Scalability and Robustness Comparison on Synthetic Large-scale Datasets. Key performance metrics are averaged over 1000 independent runs for each complexity group.}
		\label{tab:scalability}
		\setlength{\tabcolsep}{3.5pt} 
		
		\begin{tabular}{@{}lcccc@{}} 
			\toprule
			\textbf{\begin{tabular}{@{}c@{}}Complexity \\ (CC, Nodes)\end{tabular}} & 
			\textbf{Method} & 
			\textbf{\begin{tabular}{@{}c@{}}Success \\ Rate (\%)\end{tabular}} & 
			\textbf{\begin{tabular}{@{}c@{}}Coverage \\ Rate (\%)\end{tabular}} & 
			\textbf{\begin{tabular}{@{}c@{}}Time \\ (s)\end{tabular}}  \\ 
			\midrule
			
			\multirow{4}{*}{10, 9} &  BFS & 27.8 & 89.05 & \textbf{0.0000}  \\ 
			& Holistic MIP & 100 & 100 & 3.4802  \\ 
			& Incr. MIP1 & 97.5 & 99.75 & 0.1088 \\ 
			& Incr. MIP2 & 100 & 100 & 0.1163 \\ 
			\midrule 
			
			\multirow{4}{*}{50, 30} &  BFS & 17.5 & 96.52 & 0.0007  \\ 
			& Holistic MIP & 100 & $-$ & $-$  \\ 
			& Incr. MIP1 & 55.8 & 98.87 & 4.2715 \\ 
			& Incr. MIP2 & 100 & 100 & 4.6156 \\ 
			\midrule
			
			\multirow{4}{*}{100, 50} &  BFS & 12.3 & 97.94 & \textbf{0.0023} \\ 
			& Holistic MIP & 100 & $-$ & $-$ \\ 
			& Incr. MIP1 & 14.7 & 98.08 & 15.3221  \\
			& Incr. MIP2 & 100 & 100 & 17.7642 \\ 
			\bottomrule          
		\end{tabular}
	\end{table}
	\section{Discussion}
	The empirical results presented in this work signal a paradigm shift: reframing basis path generation from a procedural traversal task to a declarative optimization problem. By encoding the properties of an optimal basis set into mathematical constraints, our focus moves from \textit{how} to find paths to \textit{what} constitutes a high-quality set. This section discusses the efficacy of our incremental approach and directly addresses the critical challenge of semantic feasibility.
	
	\subsection{Efficacy of the Novelty-Driven Incremental MIP}
	A key finding of this work is the remarkable robustness of the Incremental MIP strategy when equipped with our multi-objective function (Incr. MIP2). The "Greedy Trap," where a myopic focus on path length leads to an inability to form a complete basis, is a fundamental flaw in sequential generation. Our "novelty penalty" mechanism directly counteracts this. By assigning an additional cost to the consumption of new edges, the model is mathematically incentivized to be structurally conservative, preserving uncovered edges for as long as possible. This strategic reservation of structural diversity is the primary reason for its 100\% success rate on complex topologies where all other sequential methods, including the greedy MIP variant, fail. While we do not yet offer a formal proof of equivalence, the consistent empirical convergence to a complete basis set suggests that this heuristic effectively guides the local search towards a globally viable solution.
	
	\subsection{The Infeasible Path Problem: A Bridge to Semantic Testing}
	A crucial limitation of all purely structural testing techniques is the Infeasible Path Problem \citep{Ngo2008,yao2006study}. It is essential to distinguish this \textit{semantic} challenge from the foundational structural problem of generating a complete basis set—a critical prerequisite that this paper addresses. Unlike traditional greedy algorithms, which can fail to produce a complete set and thus provide a flawed foundation for analysis, our MIP-based approach guarantees a complete and structurally minimal basis path set. This reliable "scaffold" serves as a superior input for downstream semantic tools, such as SMT solvers \citep{DeMoura2008}, which can then more tractably analyze these simpler paths to generate feasible test cases. While full integration is future work, this positions our method as a vital first step in a powerful workflow that bridges structural optimization with semantic validation, for instance, through a feedback loop where infeasibility findings refine the MIP model.

	\section{Conclusion}
	The traditional procedural approach to basis path testing was reframed in this paper as a declarative optimization problem. Instead of telling how to find paths, focusing on the definition of an optimal path set, our Mixed Integer Programming (MIP) framework provides a robust and mathematically grounded solution to mitigate the limitations of greedy algorithms. As we empirically validated, the proposed Incremental MIP strategy based on a novel multi-objective function proves to be extremely robust. It reached a 100\% success rate in producing complete basis sets over diverse and complex topologies – which are not possible with conventional methods. This proves its ability to create test path sets that are both complete and structurally minimal, reducing the cognitive overhead and delivering a better quality input source for downstream test data generation tools. At its core, the work shows that selection of basis paths is best addressed as an optimization problem. Although the problem of semantic feasibility remains, our method serves as a foundation for development. We will consider combining our MIP framework and semantic analysis engines, also known as SMT solvers, in our future work in order to develop a hybrid approach for creating path sets that are not only structurally optimal but are also guaranteed to be executable.

	\section*{Acknowledgments}
	This work was supported in part by the National Natural Science Foundation of China under Grant 62302154; in part by the Doctoral Scientific Research Foundation of Hubei University of Technology under Grants
	XJ2022007201 and XJ2022006701.

	\bibliographystyle{unsrtnat}
	\bibliography{references}

\end{document}